# Studies on the Magnetic Ground State of a Spin Möbius Strip

Graham N. Newton,[a,b] Norihisa Hoshino,[a] Takuto Matsumoto,[a] Takuya Shiga,[a] Motohiro Nakano,[c] Hiroyuki Nojiri,[d] Wolfgang Wernsdorfer,[e] Yuji Furukawa,[f] and Hiroki Oshio*[a]

Dedication

**Abstract:** Here we report the synthesis, structure and detailed characterisation of three *n*-membered oxovanadium rings, $Na_n[(V=O)_nNa_n(H_2O)_n(\alpha, \beta, or \gamma\text{-CD})_2]\cdot mH_2O$ ($n$ = 6, 7, or 8), prepared by the reactions of $(V=O)SO_4 \cdot xH_2O$ with $\alpha$, $\beta$, or $\gamma$-cyclodextrins (CDs) and NaOH in water. Their alternating heterometallic vanadium/sodium cyclic core structures were sandwiched between two CD moieties such that O-Na-O groups separated neighbouring vanadyl ions. Antiferromagnetic interactions between the $S = ½$ vanadyl ions led to $S = 0$ ground states for the even-membered rings, but to two quasi-degenerate $S = ½$ states for the spin-frustrated heptanuclear cluster.

## Introduction

There are numerous examples throughout science of systems in which their properties are dependent upon their odd/even membered nature. For example, in organic chemistry, the isotropic–nematic transition temperatures in *p*-alkoxyazoxybenzenes oscillate as their number of constituent carbon atoms increases,[1,2] and in nuclear physics, the energy term in the liquid drop model depends upon the odd/even nature of the number of nucleons.[3] Likewise in magnetism, the number of constituents has a direct effect on the physical properties. For instance, the Haldane gap, an energy gap above the ground state in antiferromagnetic Heisenberg chains, is only observed in integer (i.e. even electron) spin chains,[4,5] and spin gaps are observed in spin-ladder systems with even numbers of legs.[6,7]

Paramagnetic ring-shaped molecules have been attracting intense research interest due to their magnetic properties that may give rise to new physical phenomena as well as many exciting applications in molecular and nanomagnetism. Rings with intra-ring ferromagnetic interactions can exhibit high spin ground states and superparamagnetic behavior, with some acting as single-molecule magnets (SMMs).[8–10] Rings with intra-ring antiferromagnetic interactions exhibit magnetism dependent upon the number of constituent metal ions. Even-membered antiferromagnetic rings, in which neighbouring spins have opposite orientation (up and down spins), have singlet ground states and may exhibit quantum rotation of the Néel vector.[11] In odd-membered rings, on the other hand, all spins cannot be aligned in an alternating fashion and the residual spin can have any orientation, resulting in a spin frustrated system. An equilateral odd-membered ring composed of spin-doublet metal ions has doubly degenerate Kramers doublets (a total degeneracy of four) as the ground state.[12,13] Structural distortion can cause the doublets to split, and involving the anisotropy allows the observation of tunnelling transitions between the sublevels. As antiferromagnetic odd-membered rings have a periodic spin behaviour involving spin phase inversion, they have been described as 'quantum spin Möbius strips'.[14]

It was proposed by Loss et al. that antiferromagnetic odd-membered rings may exhibit long decoherence times,[15,16] with their ground state doublets making them excellent candidates for application as a quantum bits (qubits) in quantum information processing.[17-19] However, despite the potential for odd-membered cyclic compounds to exhibit interesting physical properties, equilateral systems containing more than three metal centres remain very rare, likely due to a combination of kinetic and thermodynamic issues as well as the lack of crystallographic odd number-fold rotational symmetries higher than three. The prime exceptions are the metallocrowns of Pecoraro et al,[20] and the family of Cr rings elegantly studied by Winpenny and co-workers,[21] who recently extended their work to introduce a new classification system for spin frustration in molecular magnets.[22]

We set out to solve this problem by using a well-defined β-cyclodextrin (CD) template. CDs are cyclic oligosaccharides, and α, β, and γ-CD molecules, which contain six, seven and eight glucopyranose groups, respectively, can be obtained from cornstarch.[23] Klüfers et al. reported an oxovanadium complex, $Na_6[(VO)_6Na_6(H_2O)_6(\alpha\text{-CDH}_{-12})_2]\cdot 59H_2O$, in which two highly-deprotonated α-CD molecules sandwich six oxovanadium ions.[24] Building on this result, we used the templating nature of β-CD to support an extremely rare example of a regular odd-membered ring spin-system in which rings of O-Na-O-bridged vanadyl ions were stabilized.[25,26] Our extended studies into the structures and magnetic properties of the β-CD complexes are presented

[a] Dr Graham N. Newton, Dr Norihisa Hoshino, Dr Takuto Matsumoto, Dr Takuya Shiga, Prof. Hiroki Oshio
Graduate School of Pure and Applied Sciences
University of Tsukuba, Tennodai 1-1-1
Tsukuba, 305-8571, Japan
E-mail: oshio@chem.tsukuba.ac.jp
[b] Dr Graham N. Newton
School of Chemistry
University of Nottingham
Nottingham, NG7 2RD, UK
[c] Prof Motohiro Nakano
Research Center for Structural Thermodynamics
Graduate School of Science, Osaka University
Machikaneyama 1-1, Toyonaka, Osaka 560-0043, Japan
[d] Prof Hiroyuki Nojiri
Institute of Materials Research
Tohoku University, Katahira 2-1-1
Aoba-ku, Sendai 980-8577, Japan
[e] Prof Wolfgang Wernsdorfer
CNRS, Institut NEEL & Univ. Grenoble Alpes
F-38000 Grenoble, France
[f] Prof. Yuji Furukawa
Ames Laboratory and Department of Physics and Astronomy
Iowa State University, Ames, IA 50011, USA

Supporting information for this article is given via a link at the end of the document.

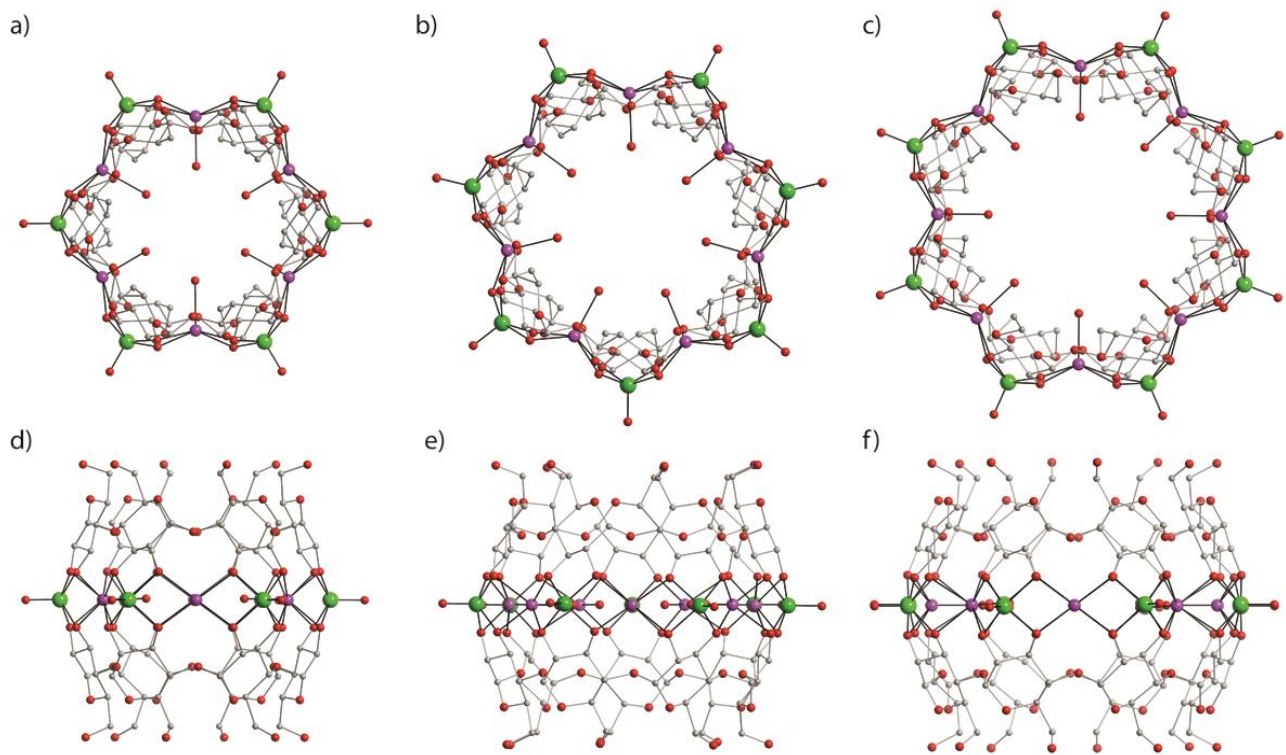

**Figure 1.** The crystal structures of complexes **1**, **2**, and **3**. a) **1** (top view), b) **2** (top view), c) **3** (top view), d) **1** (side view), e) **2** (side view), f) **3** (side view). Colour code: vanadium, green; sodium, pink; carbon, grey; oxygen, red. Hydrogen atoms, counter cations and lattice solvent molecules are omitted for clarity

herein, and compared to those of the related hexa- and octanuclear complexes templated by α-CD, and γ-CD ligands, respectively. Detailed magnetic analyses were conducted on all samples, and their spin ground states investigated using micro-SQUID and solid-state NMR techniques.

## Results and Discussion

The reactions of $(V=O)SO_4 \cdot nH_2O$ with α, β and γ-CD and NaOH in water yielded green solutions, the diffusion of acetone into which gave blue hexagonal rods of $Na_6[(V=O)_6Na_6(H_2O)_6(\alpha\text{-CD})_2]\cdot nH_2O$ (**1**), $Na_7[(V=O)_7Na_7(H_2O)_7(\beta\text{-CD})_2]\cdot nH_2O$ (**2**), and $Na_8[(V=O)_8Na_8(H_2O)_8(\gamma\text{-CD})_2]\cdot nH_2O$ (**3**), respectively, all of which lost solvent when dried in air, with crystals of **2** exhibiting the most rapid loss.

**Structural studies:** Single crystal X-ray structural analyses were conducted for **1** - **3** (Figure 1) and crystallographic parameters, selected bond distances and angles are shown in the Supplementary Information. Structure **1** was described by Klüfers et al. in the $P312$ space group,[24] however in the present study the core structure of **1** was solved in the $P622$ space group and the disordered lattice water molecules were treated with the PLATON/SQUEEZE program.[27] In attempts to lower the symmetry of the structural solution, the $C_6$ symmetry in the core was always retained, while hydrogen bonded networks consisting of $Na^+$ and $H_2O$ molecules located between the clusters were not modelled satisfactorily even in the $P1$ space group. Crystal structures of **2** and **3** were dealt with in the same manner as **1**. 993, 724, and 1360 electrons per core complex were observed by PLATON/SQUEEZE for **1**, **2** and **3** respectively, corresponding to 6 $Na^+$ and 94 $H_2O$ molecules (1000 electrons), 7 $Na^+$ and 65 $H_2O$ molecules (720 electrons), and 8 $Na^+$ and 128 $H_2O$ molecules (1359 electrons) respectively. Such assumptions were reasonable for the observed void volumes (2217, 2802, and 3401 Å$^3$ per cluster in crystals of **1**, **2** and **3** respectively), however deviated from the formulae calculated by elemental analyses due to solvent loss and/or the existence of $[Na^+_a(H_2O)_b(OH^-)_c]^{a-c}$ clusters. Hence, precise numbers of the solvent molecules in the crystals were unclear. It should be noted, however, that the formulae estimated from the X-ray data were in good agreement with those implied by magnetic analyses.

The structures of all compounds are very similar, consisting of a ring of vanadyl ions separated by O-Na-O bridges, sandwiched by two CD units of which all secondary alcohol groups are deprotonated. Glucopyranoses are so rigid that CDs have pseudo-rotational symmetries, which are transferred to the vanadyl ions of **1**-**3**. **1** has ideal $C_6$ symmetry rendering all six of its vanadyl ions crystallographically equivalent, while **2** has a two-fold axis[25] and **3** has two mutually-crossing $C_2$ axes in the ring plane.

The neighbouring vanadium ions in the rings are separated by O-Na-O bridges with interatomic distances of 6.449(9) Å in **1**, 6.197(2) – 6.396(2) Å in **2** and 6.246(1) – 6.388(1) Å in **3**. The interior angles of the vanadyl rings are 120° in **1**, 125.76(3) – 130.10(3)° in **2** and 133.52(1) and 136.47(1)° in **3**, values close to those of ideal hexagons (120°) heptagons (128.57°) and octagons (135°), respectively. The vanadium ions have square pyramidal coordination geometries, where the basal sites are coordinated by oxygen atoms from secondary alcohols with bond lengths of V-O = 1.949(3) – 1.986(3) Å for **1**, 1.915(4) – 2.013(4) Å for **2**, and 1.935(3) – 1.999(3) Å for **3**. The axial sites are occupied by oxo ions with bond lengths of 1.634(3) Å for **1**, 1.575(6) – 1.629(6) Å for **2** and 1.619(2) and 1.650(2) Å for **3**. The sodium ions also have square pyramidal coordination geometries, where four alkoxo-oxygen atoms from the CD groups occupy the basal positions and a water molecule coordinates in the axial site.

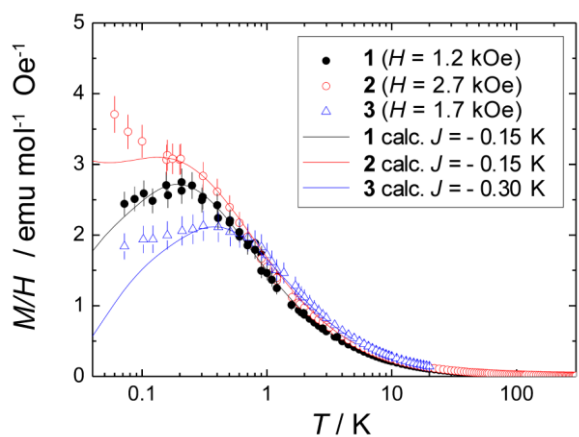

**Figure 2.** The temperature dependence of *M*/*H* for **1** (black), **2** (red) and **3** (blue). Calculated values are shown as solid curves. The values below 1.8 K were estimated from the FWHA data.

**Magnetic studies:** Magnetic susceptibility data were collected on fresh microcrystalline samples of **1**, **2** and **3** in the temperature range of 1.8 – 300 K in an applied field of $H$ = 10 kOe (Figure S1). Magnetic susceptibility data above 1.8 K obeyed the Curie-Weiss law with Curie and Weiss constants of $C$ = 2.17 cm$^3$ mol$^{-1}$ K, $\theta$ = -1.03 K; $C$ = 2.75 cm$^3$ mol$^{-1}$ K and $\theta$ = -1.02 K; and $C$ = 3.09 cm$^3$ mol$^{-1}$ K, $\theta$ = -0.66 K for **1**, **2** and **3** respectively, using the number of solvent molecules obtained from the crystallographic studies. The values of **2** and **3** were slightly larger than the expected values from HF-EPR data (see below). The numbers of the $H_2O$ molecules were re-estimated as **1**·94$H_2O$, **2**·85$H_2O$ and **3**·110$H_2O$ using the HF-EPR data ($C$ = 2.18, 2.36 and 2.91 cm$^3$ mol$^{-1}$ K for **1**, **2** and **3**) and corrected $\chi_m T$ versus $T$ plots were depicted in Figure S2. All data sets showed the $\chi_m T$ values to remain constant between 300 and 30 K, below which they decreased monotonically as temperature was lowered, indicating that antiferromagnetic interactions were operative between neighbouring vanadyl centres in all samples.

Single crystal HF-EPR spectra for **1**, **2** and **3** showed single isotropic peaks of $g$ = 1.970, 1.973 and 1.968, corresponding to the doublet spins of $VO^{2+}$ ions (Figure S3). No temperature dependence was observed in the range of 4.2-20 K.

As a means of interrogating the spin ground state of the three compounds, solid-state $^1H$ nuclear magnetic resonance (NMR) measurements were conducted on **1**, **2** and **3** down to 60 mK (Figure 2, S4) using deuterated (with the exception of the CD alkyl protons) samples that were shown from X-ray structural analysis to be isomorphous to the non-deuterated crystals. The linewidths (full width at half amplitude, FWHA) for **1** and **3** increase with decreasing temperature $T$, reaching broad peaks around 0.3 – 0.4 K before decreasing at lower temperatures. In contrast, the linewidths for **2** continued to increase as the temperature was lowered to 60 mK. The linewidth can be expressed in terms of magnetization, $M$, as FWHA = $a + bM$ where $a$ is the $T$- and $H$-independent linewidth due to nuclear dipolar interactions, and $b$ is related to the hyperfine coupling constants of $^1H$. Using the magnetization data above 1.8 K collected by SQUID magnetometer, $a$ and $b$ were estimated to be 4.3 Oe and 6.5 Oe mol emu$^{-1}$, 4.8 Oe and 9.0 Oe mol emu$^{-1}$ and 4.0 Oe and 5.2 Oe mol emu$^{-1}$ for **1**, **2**, and **3**, respectively. The temperature dependence of *M*/*H* for the three crystal samples below 1.8 K was then estimated from the FWHA data. The obtained *M*/*H* plots mirrored the FWHA data, with the broad peaks and subsequent decreases observed for **1** and **3** indicative of spin singlet $S$ = 0 ground states. In contrast, the unsaturated increase of *M*/*H* for **2** at low temperatures clearly demonstrates that the ground state is not spin singlet but magnetic, $S \neq 0$. These results are consistent with the micro-SQUID results shown below.

The *M*/*H* data collected for the rings were fitted using the following simple Heisenberg spin Hamiltonian:

$$\hat{H} = -2\sum_{i=1}^{m-1}(J_i\hat{S}_i \cdot \hat{S}_{i+1}) - 2J_m\hat{S}_m \cdot \hat{S}_1 - g\beta\mu_0 H\sum_{i=1}^{m}\hat{S}_i \quad (1)$$

where $J_i$ is the nearest-neighbour superexchange interaction parameter between the $i$-th and ($i$ + 1)-th vanadyl ions, $m$ is the number of constituent vanadium centres, $g$ is the Landé $g$ factor, $\beta$ is the Bohr magneton, $\mu_0$ is the permeability of vacuum, and $H$ is external magnetic field. Single-$J$ approximation ($J_i \equiv J$) gives a negative exchange constant, $J$, of which the magnitude was less than 1 K. Solid curves in Figure 2 show the fitting results of *M*/*H* using (*M*/*H*)$_{cal}$ calculated at corresponding magnetic fields for the three vanadium rings, where a molecular field parameter, $\lambda$ = −0.1 – −0.15Oe mol emu$^{-1}$ defined as (*M*/*H*)$^{-1}$ = (*M*/*H*)$_{cal}^{-1}$ − $\lambda$, was introduced to account for the intermolecular magnetic interactions. From the fittings, small antiferromagnetic exchange interactions ($J$) were roughly estimated to be −0.15 – −0.30 K for the three vanadium rings. The deviation between the experimental and calculated *M*/*H* at very low temperatures below 0.1 K could be due to effects of nuclear hyperfine field on $^{51}V$ as will be discussed below.

In order to obtain information on the detailed spin states of the rings, low temperature magnetization experiments were carried out on single crystals at 0.04 – 0.9 K using a micro-

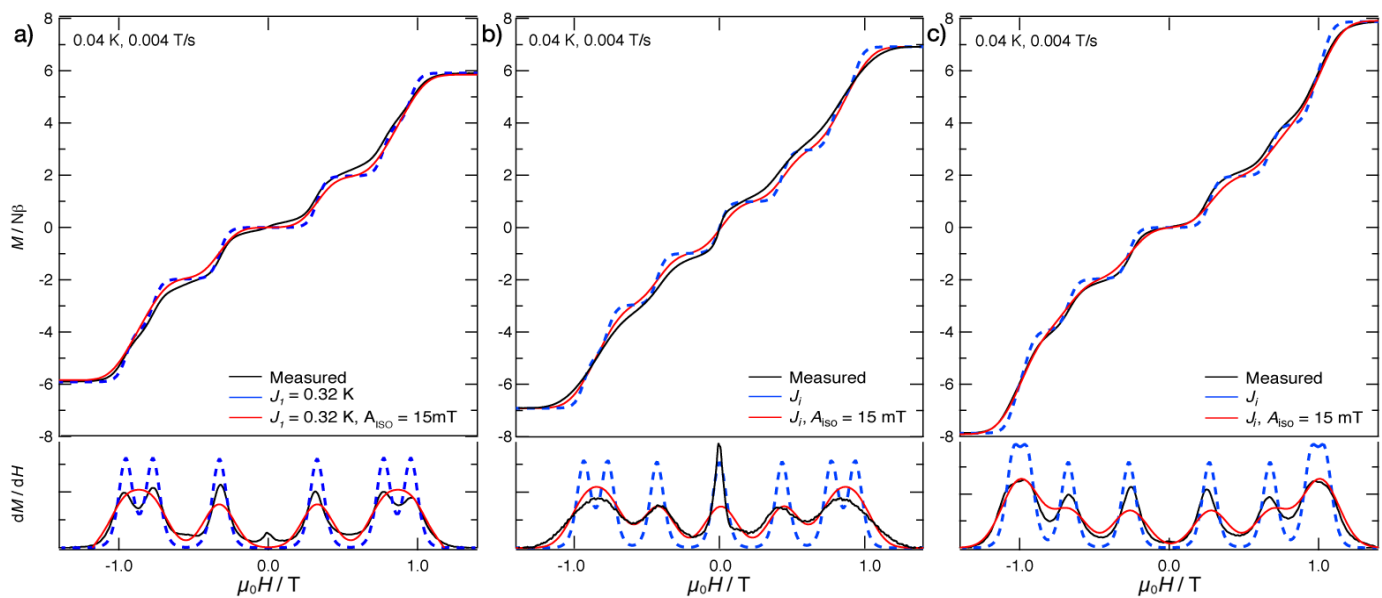

**Figure 3.** Magnetization vs. external magnetic field, and field derivative plots collected at 0.04 K at a sweep rate of 0.004 T/s for complexes **1** (a), **2** (b) and **3** (c). The black curves represent the observed data, while the blue and red plots are the fitting simulations calculated using only the $J_i$ values, and using the $J_i$ values in conjunction with the hyperfine field of the vanadium nuclei, respectively.

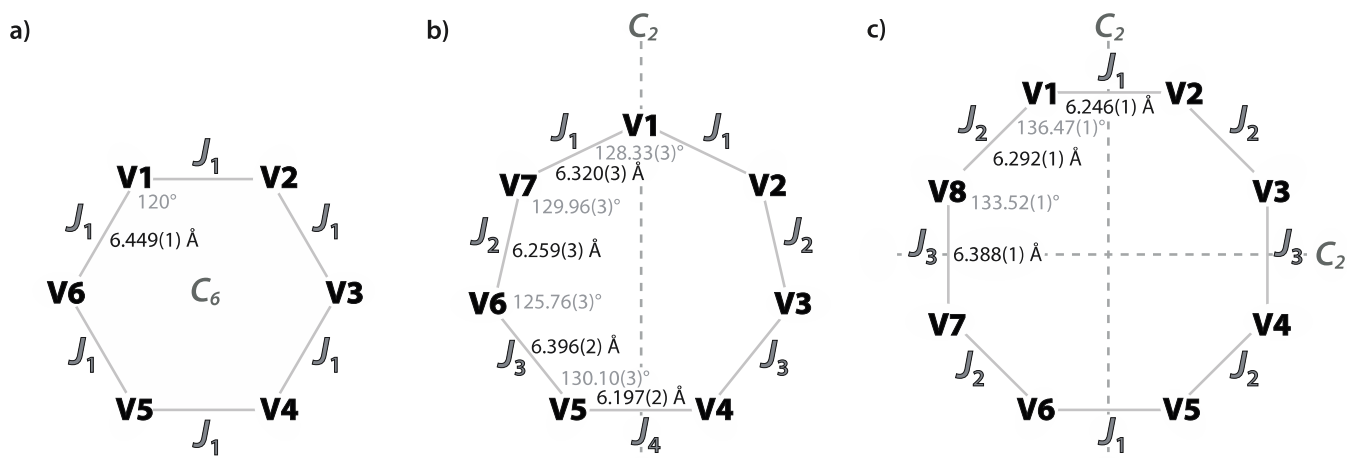

**Figure 4.** Schematic core structures of **1** (a), **2** (b) and **3** (c) indicating symmetry operators and the magnetic coupling pathways used in the fitting calculations.

SQUID magnetometer. At temperatures higher than 0.25 K, the $M/M_S$ values for all samples increased monotonically as the magnetic field was increased, however, the $M/M_S$ versus $\mu_0 H$ plots clearly showed several plateaus below 0.1 K (Figures S5-7). $M$ versus $\mu_0 H$ plots collected at 0.04 K at a scan rate of 0.004 Ts$^{-1}$ were obtained using the HF-EPR data ($M_S = gS$ = 5.916, 6.901 and 7.872 for **1**, **2** and **3**), and are depicted in Figure 3 with corresponding field derivative plots. The results clearly showed the presence of ground spin-level crossings at several magnetic fields induced by the Zeeman effect and the stepped-shapes reflect the different antiferromagnetic interactions in **1**, **2** and **3**. It should be noted that the exchange coupling constants for the rings were small enough to cover all magnetization steps within 1.5 T. The crossing fields provide useful information about intra-ring exchange interactions between vanadium ions. The vanadium ions have $S = 1/2$ spin states and are considered to be magnetically isotropic. The magnetization data were, therefore, analysed using the same spin Hamiltonian used for the analysis of $^1$H NMR results.

The $C_6$ symmetry of complex **1** ensures that six exchange paths should be represented with a single exchange parameter $J$ (Figure 4), while the crystallographic $C_2$ axis of **2** means that four exchange parameters, $J_1$ (= $J_7$), $J_2$ (= $J_6$), $J_3$ (= $J_5$) and $J_4$ were required.[26] The two $C_2$ axes of complex **3** ensure that three exchange parameters are sufficient. Least-squares calculations for the magnetization curves yielded $J$ = -0.32 K ($g$

= 1.97) for **1**, $J_1$ = -0.31 K, $J_2$ = -0.26 K, $J_3$ = -0.17 K, and $J_4$ = -0.33 K (*g* = 1.97) for **2**, and $J_1$ = -0.32 K, $J_2$ = -0.34 K, and $J_3$ = -0.38 K (*g* = 1.97) for **3**. The *g* values used for the calculation were obtained by single crystal EPR measurements at 4.2 K (Figure S3).

The magnetization curves calculated using the obtained $J_i$ values reproduced the numbers of level-crossing fields of the lowest Zeeman split sub-levels ($m_s$) expected for the 6-, 7- and 8-membered rings; **1** showed sub-levels corresponding to *S* = 1, 2 and 3, **2** showed four spin states corresponding to *S* = 1/2, 3/2, 5/2, and 7/2, and **3** showed *S* = 1, 2, 3 and 4. However, the observed curves were broader than those simulated (blue curves in Figure 2 (bottom)), and the implementation of anisotropic exchange coupling constants into the calculation did not improve the fits. A linewidth function to account for the nuclear hyperfine field ($A_{iso}$ = 15 mT for each vanadium ion (*I* = 7/2 for $^{51}$V; natural abundance is 99.75%)) was then considered,[28–30] and the resultant fitting curves accurately reproduced the observed d*M*/d*H* versus $\mu_0H$ plots for all samples (the red curves in Figure 3).

**Distortion effects on the magnetic properties of odd-membered rings:** Replacement of the diffusion solvent (acetone) used in the crystallization of **2** with methanol allowed isolation of blue plate crystals of Na$_7$[(V=O)$_7$Na$_7$(H$_2$O)$_7$(β-CD)$_2$]·*n*H$_2$O (**2'**), a second polymorph of the seven-membered ring (Figure 5, Tables S1&2). If the vertices of a 7-sided polygon are co-planar, the total interior angle should be 900°. The corresponding angles of 900.1° and 899.9° for **2** and **2'**, respectively, indicate that both rings can be considered planar heptagons, however, subtle ring distortions in their structures have major impacts on their magnetic properties. While **2** has a crystallographic $C_2$ axis, and the magnetic coupling can therefore be effectively approximated by four exchange parameters, every vanadyl ion in **2'** is crystallographically independent. Despite this, we felt that similarities in inter-metallic distances of some of the vanadyl pairs meant that the model could be similarly simplified to four exchange parameters: $J_1$ (= $J_3$), $J_2$ (= $J_7$), $J_4$ and $J_5$ (= $J_6$) (Figure 5).

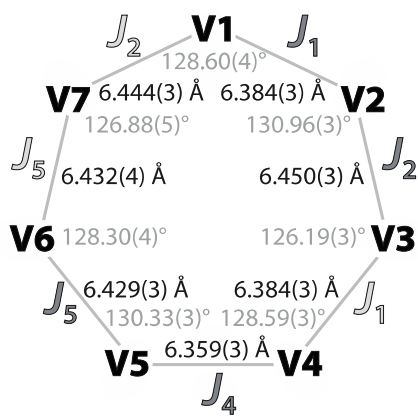

**Figure 5.** Schematic diagram of the core structure **2'**, and of the magnetic coupling parameters used in the data fitting. *J* parameters in light grey are approximated based on bond distances.

Field dependent *M* versus $\mu_0H$ plots for complex **2'**, collected at 0.04 – 0.90 K at a scan rate of 0.004 Ts$^{-1}$, were analysed using the spin Hamiltonian of eq. 1 to yield exchange parameters of $J_1$ = -0.32 K, $J_2$ = -0.16 K, $J_4$ = -0.30 K, and $J_5$ = -0.60 K for (*g* = 1.98) (Figure 6).

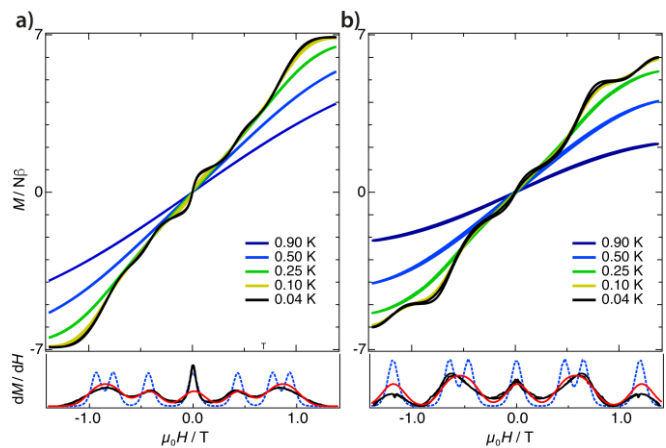

**Figure 6.** Magnetization (at 0.04 – 0.90 K) and field derivative versus external field plots at 0.04 K for (a) **2** (partially reproduced from Figure 3b for direct comparison) and (b) **2'**, respectively. As in Figure 3, the black, blue dotted, and red curves in the d*M*/d*H* versus $\mu_0H$ plots correspond to the observed data, the simulation curves with the $J_i$ values, and the curves simulated with the $J_i$ values and the hyperfine field of the vanadium ions, respectively.

Spin chirality renders the ground states (*S* = 1/2) of the antiferromagnetic odd-membered rings doubly degenerate.[12,13,31,32] In the classical model, the two states have spin alignments with each spin tilted with an angle of +128.57° or -128.57° (= the interior angle of a regular heptagon), corresponding to clockwise and counter clockwise orientations, respectively. Kouzoudis proposed an approach to allow the effective estimation of the ground state.[33] Degeneracy will be lifted by the distortion of the heptagon, leading to two split Kramers doublets with spin quantum numbers of $S_1$ = 1/2 and $S_2$ = 1/2 (Figure 7). The Zeeman splitting scheme for these doublets has four level crossings, in which small anisotropic interactions such as the Dzyaloshinsky-Moriya (DM) interactions[34,35] lead to avoided-crossings at non-zero crossing fields. Landau-Zener-Stückelberg (LZS) type transitions,[36–38] which are due to DM interactions or transverse magnetic field, are expected at the avoided-crossings. Two asymmetric peaks in the d*M*/d*H* versus *H* plots of **2** and **2'** were observed around $\mu_0H$ = 0 T (Figure 8). It can be considered that the observed peak splitting is due to spin flips between the sublevels ($m_s$ = ±1/2) of the non-degenerate Kramers doublets. Supposing the peak maxima correspond to the level-crossing fields of the Zeeman split sublevels, the peak separations (0.01 and 0.02 T for **2** and **2'** at 4 mT/s) correspond to energy gaps of 2.7 and 5.4 mK between $S_1$ and $S_2$ states for **2** and **2'**, respectively. The larger energy gap for **2'** is in good agreement with its more distorted structure.

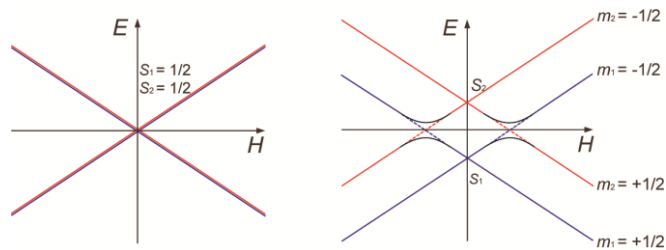

**Figure 7.** Zeeman splitting diagrams of (left) doubly degenerate and (right) non-degenerate spin states for regular and distorted heptagons, respectively.

Magnetization curves are usually symmetric to the magnetic field, however, the peaks around $\mu_0 H = 0$ T in the d$M$/d$H$ versus $\mu_0 H$ plots were smaller in the positive field than those in the negative field when the magnetic field was swept from the positive field, and *vice versa* (Figure 8). The observed peak asymmetry may arise from different thermal populations of the two Kramers doublets (Figure 7). When the magnetic field was swept from the positive field and reached the first level-crossing point, the up-spin on the upper sublevel ($S_2$, $m_2$ = +1/2) flips to the ($S_1$, $m_1$ = -1/2) state through an LZS type transition, while the spin on the lowest sublevel ($S_1$, $m_1$ = +1/2) does not flip at the same field. Under a negative field, the residual up-spin ($S_1$, $m_1$ = +1/2) flipped to the down spin ($S_2$, $m_2$ = -1/2). The ratios of thermal population on the $S_1$ and $S_2$ states were 0.94 and 0.85 at 40 mK for **2** and **2'**, respectively, estimated by the energy gaps between the $S_1$ and $S_2$ states. Although our experiments could conceivably be in the phonon bottleneck and the estimation may not be so accurate, such disparity should lead to the observation of asymmetric peaks. It is also interesting to note that this asymmetry is an indication that the system is not in thermal equilibrium.

The effects of field sweep rate on the differential magnetization plots were also examined, and found to exhibit contrasting behaviour between **2** and **2'** (Figure 8). The higher symmetry ring, **2**, showed coalescence of the apparent crossing field with increasing sweep rate, while the two-peak profile of the lower symmetry ring, **2'** was essentially sweep-rate independent. The former observation can be attributed to the transition probabilities in non-zero crossing fields being very small and therefore ensuring that sweep rate dependence is observed in the range of 4-280 mT/s. In the more distorted molecule, the degree of level mixing is far greater, and the transition is in the adiabatic regime and independent of the sweep rate in the measurement. Note that the sweep rate dependence in Fig.8 is quite different from the phonon bottleneck in the previous reports of V$_{15}$ and general LZS systems.[39,40]

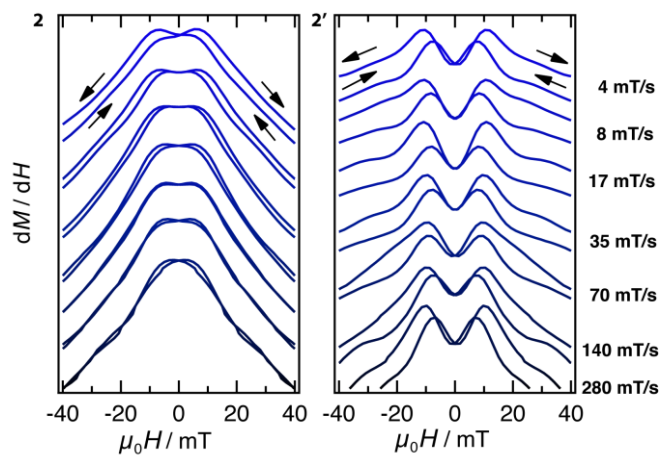

**Figure 8.** d$M$/d$H$ versus $\mu_0 H$ plots of **2** (left) and **2'** (right) for several sweep rates (4 – 280 mT/sec ) at 0.04 K.

## Conclusions

In conclusion, a family of even- and odd-membered spin rings was synthesized and characterized both structurally and magnetically. The odd-membered rings were extremely rare examples of homospin seven-membered rings, and their magnetism was studied in detail to probe the nature of their spin coupling. Solid-state NMR studies conducted down to 50 mK indicated that the even-membered rings exhibited spin singlet $S$ = 0 ground states, while the odd-membered species had a magnetic ground state. Fitting of the magnetization data collected at 40 mK allowed the magnitude of the exchange couplings between neighbouring spins to be deduced.

A second heptagonal species of lower symmetry was isolated allowing investigation of the effects of structural distortion on the low temperature magnetic behaviour. The heptagons have two Kramers doublets as the spin ground state, and the energy splitting depends upon ring distortions. Magnetization experiments at 40 mK showed stepped magnetization curves, characteristic of level crossings between spin sublevels, and suggested that the quantum tunnelling dynamics of the spin flips depend upon the distortion of the rings.

## Experimental Section

**Synthesis of Na$_6$[(V=O)$_6$Na$_6$(H$_2$O)$_6$($\alpha$-CD)$_2$]·$n$H$_2$O (1).** Oxovanadium(IV) sulphate hydrate (55 mg, 0.25 mmol) and $\alpha$-CD (70 mg, 0.062 mmol) were suspended in water (1.0 ml). Sodium hydroxide (55 mg, 1.34 mmol) and $\alpha$-CD (70 mg, 0.062 mmol) were dissolved in water (0.5 ml) and added to the suspension. After stirring for several minutes, the solid oxovanadium(IV) sulphate dissolved to yield a green solution. Vapour diffusion of acetone in a sealed tube gave blue hexagonal rod-like crystals. Anal. calcd. for C$_{72}$H$_{228}$Na$_{12}$O$_{132}$V$_6$ (**1**·**60H$_2$O**): C 22.83, H 6.07; found: C 22.79, H 5.02.

**Synthesis of Na$_7$[(V=O)$_7$Na$_7$(H$_2$O)$_7$($\beta$-CD)$_2$]·$n$H$_2$O (2, 2').** Oxovanadium(IV) sulphate hydrate (55 mg, 0.25 mmol) and $\beta$-CD (70 mg,

0.062 mmol) were suspended in water (0.5 ml). Sodium hydroxide (128 mg, 3.2 mmol) and β-CD (70 mg, 0.062 mmol) were dissolved in water (0.5 ml) and added to the suspension. A dark green solution was yielded and vapour diffusion of acetone and methanol in sealed tubes gave blue hexagonal rod-like crystals of **2** and blue prismatic crystals of **2'**, respectively. Anal. calcd. for $C_{84}H_{306}Na_{14}O_{174}V_7$ (**2·90H$_2$O**): C 21.11, H 6.45; found: C 21.43, H 5.79. Anal. calcd. for $C_{84}H_{224}Na_{14}O_{133}V_7$ (**2'·49H$_2$O**): C 26.52, H, 5.24; found C 25.01, H 5.36.

**Synthesis of Na$_8$[(V=O)$_8$Na$_8$(H$_2$O)$_8$(γ-CD)$_2$]·nH$_2$O (3)**. Preparation of **3** was achieved using the same synthetic approach employed in the isolation of **1**, using γ-CD as the capping ligand. Hexagonal plate-like crystals were obtained. Anal. calcd. for $C_{96}H_{304}Na_{16}O_{176}V_8$ (**3·80H$_2$O**): C 22.83, H 6.07; found C 22.77, H 4.77.

**Crystallography.** Each single crystal was mounted on the tip of a glass fibre with epoxy resin. Diffraction data were collected at −73 °C on a Bruker SMART APEX diffractometer fitted with a CCD-type area detector using monochromated Mo-K$\alpha$ radiation ($\lambda$ = 0.71073 Å). The data frames were integrated using SAINT (Bruker Analytical X-ray Systems). Gaussian face-indexed corrections were applied by XPREP (Bruker Analytical X-ray Systems). The structures were solved by direct methods and refined by the full-matrix least-squares method on all $F^2$ data using the SHELXTL program package (Bruker Analytical X-ray Systems). Non-hydrogen atoms were refined with anisotropic thermal parameters. Hydrogen atoms were included in calculated positions and refined with isotropic thermal parameters riding on those of the parent atoms. The data were treated with the PLATON/SQUEEZE program to deal with solvent molecules.[27] 993 electrons per $[(VO)_6Na_6(H_2O)_6(\alpha-CD)]^{6-}$, 724 electrons per $[(VO)_7Na_7(H_2O)_7(\beta-CD)]^{7-}$, and 1360 electrons per $[(VO)_8Na_8(H_2O)_8(\gamma-CD)]^{8-}$ were observed in the void spaces of crystals **1**, **2** and **3**, respectively. Crystallographic parameters are given in the Supplementary Information, and selected bond lengths and angles are shown in Tables S1 and S2. The crystallographic information files for **1**, **2**, **2'** and **3** can be obtained free of charge from the CCDC with requisition numbers 1453312, 730918, 730919 and 1453313.

**HF-EPR spectra.** High-Field Electron Paramagnetic Resonance (HF-EPR) measurements were performed. Each single crystal was mounted on the stage of polyethylene with a very small amount of 1,1-diphenyl-2-picrylhydrazyl (DPPH) radical and radiated with the microwave oscillating at 95 - 190 GHz using Gunn oscillators, applying a rapidly pulsed field of 0 - 6 T. Resonance magnetic fields were corrected using the DPPH radical doublet ($g$ = 2.0023193).

**Magnetic susceptibility.** Magnetic susceptibility measurements in the temperature range of 1.8–300 K were carried out on powdered samples with applying magnetic field of 1 T, using a Quantum Design MPMS XL-5 type magnetometer. Corrections are based on subtracting the sample-holder signal and the diamagnetic contribution was estimated from the Pascal constants. Low-temperature (< 1.8 K) hysteresis loops were measured at Institut Néel-CNRS (Grenoble, France) using arrays of micro-SQUID.

**Solid-state NMR.** Solid-state nuclear magnetic resonance (NMR) measurements down to 50 mK at various frequencies were carried out on $^1$H ($I$ = 1/2; $\gamma/2\pi$ = 42.5774 MHz/T) using a homemade phase coherent spin-echo pulse spectrometer. The low temperature NMR measurements were performed with an Oxford Kelvinox dilution refrigerator installed at the Ames Laboratory. NMR spectra were obtained either by Fourier transformation of the NMR echo signal or by sweeping the magnetic field $H$. The NMR echo signal was obtained by means of Hahn echo sequences with a typical π/2 pulse length of 1.5 μs.

## Acknowledgements

This work was supported by a Grant-in-Aid for Scientific Research from the Ministry of Education, Culture, Sports, Science and Technology, Japan. A part of this research was supported by the U.S. Department of Energy, Office of Basic Energy Sciences, Division of Materials Sciences and Engineering. Ames Laboratory is operated for the U.S. Department of Energy by Iowa State University under Contract No. DE-AC02-07CH11358. This work was also partly performed under the Inter-University Cooperative Research Program of the Institute for Materials Research, Tohoku University.

**Keywords:** Cluster compounds • Macrocyclic ligands • Vanadium • Cyclodextrins • Magnetic properties